\documentclass[entropy,article,accept,pdftex,oneauthor]{Definitions/mdpi} 

\firstpage{1} 
\pubvolume{27}
\issuenum{4}
\articlenumber{395}
\pubyear{2025}
\copyrightyear{2025}
\externaleditor{Andrei Khrennikov}
\datereceived{10 March 2025} 
\daterevised{2 April 2025} 
\dateaccepted{3 April 2025} 
\datepublished{7 April 2025} 
\hreflink{https://\linebreak doi.org/10.3390/e27040395} 

\Title{The Dirac Equation, Mass and Arithmetic by Permutations of Automaton States}

\TitleCitation{The Dirac Equation, Mass and Arithmetic by Permutations of Automaton States}

\Author{{Hans-Thomas} 
	Elze \orcidA{}}

\AuthorNames{Hans-Thomas Elze}

\AuthorCitation{{Elze,} 
	H.-T.}

\address[1]{%
	Dipartimento di Fisica ``Enrico Fermi'', Universit\'a di Pisa, Largo Pontecorvo 3, 56127 Pisa, Italy; elze@df.unipi.it}

\abstract{The cornerstones of the Cellular Automaton Interpretation of Quantum Mechanics are its underlying ontological states that evolve by permutations. They do not create would-be quantum mechanical superposition  states. We review this with a classical automaton consisting of an Ising 
	spin chain which is then related to the Weyl equation in the continuum 
	limit. Based on this and generalizing, we construct a new 
	``Necklace of Necklaces'' automaton with a torus-like topology that lends 
	itself to represent the Dirac equation in 1 + 1 dimensions. Special attention 
	has to be paid to its mass term, which necessitates this 
	enlarged structure and a particular scattering operator contributing to the 
	step-wise updates of the automaton. As discussed earlier, such deterministic models of discrete spins or bits unavoidably become quantum mechanical, when only \linebreak slightly deformed.} 


\begin{document}
	
	\section{Introduction} 
	
	
	This work is motivated by an hypothetical deterministic ontology 
	%
	%
	to  understand natural phenomena which 
	are studied in physics and chemistry and possibly beyond. 
	This might contrast with the perception of 
	an intrinsic fundamental randomness especially in the dynamics of 
	microscopic objects, such as Standard Model `elementary' particles, atoms, 
	or molecules, which is described, for example, in the references~\cite{GisinBook,Haag2013,Haag2016}. 
	Related ideas are obviously founded on very successful developments and applications of quantum theory in the study and experimental manipulation 
	of the microscopic world.   
	
	Here, we do {\it{not}} base our discussion on quantum theory. Yet, we will quickly find it convenient to borrow elements from its most efficient  language and formalism. Our working hypothesis is what we observe in experiments, we `see' in terms of descriptions employing the language of mathematics and especially of quantum theory. 
	
	We follow the {\it{Cellular Automaton Interpretation of Quantum Mechanics}},  namely that quantum mechanical features can be found 
	in the behavior of certain classical cellular automata~\cite{tHooftBook,tHooft2023}. Thus, deterministic dynamics is 
	considered to underlie any change occurring with an object, {i.e.,} a 
	change of its ontological state.   
	
	The universe is always in a specific, even if unknown to us, ontological state. 
	In the `toy models' presented in the following, the model universe consists of 
	a countable number of two-state Ising spins that evolve 
	according to deterministic automaton rules. Incidentally, the yes--no or spin-up--spin-down alternatives here may be described as occupation numbers of fermionic degrees of freedom, i.e., with values 0 or 1. This 
	possibility has been explored in quantum field theories of fermions built 
	from classical {\it{probabilistic cellular automata}}, {e.g., } 
	in 
	reference~\cite{WetterichDICE22}.  
	
	We shall stick to strictly deterministic evolutions of cellular automata here, as in earlier work~\cite{PRA2014}, and will not discuss anew 
	arguments for and against attempts to 
	reinstall determinism at a fundamental level of natural science. Similarly, 
	we refer the reader to the literature concerning the measurement problem of 
	quantum theory, which in an ontological theory does not exist~\cite{tHooftBook}, and the violation of Bell's inequality, which does not necessarily rule out such a theory~\cite{tHooftBook,Vervoort1,Vervoort2,Hossenfelder,Vervoort3}. 
	
	These considerations invoke, in one way or another, a 
	{\it{superdeterministic}} picture of the universe. This is most clearly 
	visible in 't~Hooft's explanation of the non-classical correlations,  experimentally observed as violations of Bell's inequality, by vacuum correlations established in  the past 
	of such an experiment~\cite{tHooftBook}. However, the notion of vacuum correlations has a precise meaning only in {\it{quantum}} many-body or field 
	theory and, therefore, it seems plausible but not convincing to invoke this for the understanding of such an essential effect from an ontological 
	perspective. There is no ontological model yet that produces deterministically such quantum fluctuations. However, as far as deterministic ontological models, such as the Cogwheel Model ({Appendix}
	~\ref{appendixa.1}), have a 
	quantum mechanical limit or become quantum mechanical ``by mistake'' (cf. below); however, they do also open an avenue to understand quantum fluctuations and correlations. Clearly, this remains to be explored and is not our aim 
	here. -- An interesting probabilistic approach is presented by Wetterich that bridges between classical determinism (such as underlying  classical statistical mechanics) and quantum mechanics (possibly arising by  evolution from certain probabilistic initial conditions for otherwise  deterministic dynamics)~\cite{WetterichI}. 
	
	Instead, we shall discuss interesting features of ontological states 
	realized in classical Ising models which evolve deterministically 
	through permutations. 
	
	We remark that there are {\it {no superposition states}} created by any dynamics based on permutations. 
	Superpositions do not exist in a set 
	of ontological states; yet, they are  conveniently introduced in our mathematical language, the {\it {quantum superpositions}} of ontological (micro) states, which do not exist `out there' as states the universe could be in. On the other hand, {\it  {classical states}} are ontological (macro) states, or probabilistic superpositions thereof, appropriate for nature's phenomena at vastly different scales~\cite{tHooftBook}. -- The question of practical as well as foundational interest, whether there  is a consistent notion of quantum-classical hybrids (cf.~\cite{Elzehybrid}), has been asked in the present context with a negative answer~\cite{ElzehybridCA}. 
	
	We do not aim to replace quantum theory as the physicist's perfect toolbox. However, a better understanding of its peculiarities is desirable, just as it is desirable to recover the structural elements of its mathematical language. 
	Our approach should be seen as an attempt in this direction, based on the 
	assumption of a deterministic ontology and realized in terms of classical 
	two-state Ising spins or bits. This is distinctly different technically and conceptually, but similar in spirit to other more or less evolved, though critically discussed approaches to ``interpret'' quantum theory. -- For example, Bohmian 
	mechanics~\cite{Bohm52,Tumulka18,Gisin21}, stochastic electrodynamics~\cite{Santos22,Boyer19,CettoEtAl15},  
	prequantum classical statistical field theory~\cite{Khrennikov15}, 
	or quantum Bayesianism ({\it {QBism}})~\cite{FuchsSchack13} all  
	study the emergence, rewriting, or generalization of quantum theory 
	from various perspectives. However, 
	none of the latter interpretations really go outside of 
	quantum theory. They are occupied with not getting into conflict with its perfectly working theoretical ``machinery'' while trying to add elements helping its understanding. -- Thus, Bohmian mechanics {\it {adds}} to the continuum theory of 
	Schr\"odinger's equation for the wave function apparently classical trajectories of point particles, which are guided by the wave function. An ontology that would complete this interpretation and allow us to understand the where-from and what-for of the wave function is 
	missing. On the other hand, {\it {QBism}} does {\it {not}} seem to address at 
	all the ``stuff'' we perceive ``out there'' and describe in physics. It rather tries to make comprehensible a central aspect of quantum theory, 
	its probabilistic character with Born's rule, by relating it to 
	rationally acting agents that 
	can make predictions, verify them, and improve prior beliefs. -- These  interpretations, and likewise others not mentioned 
	here, seem far from the program advocated by 't\,Hooft and followed in a specific way in this article. 
	
	Before introducing more detailed models in this paper, it may be 
	worthwhile to recall that models based on permutations of ontological 
	states lead to Hamiltonians which are rather poor in tunable parameters 
	or coupling constants. In the long run of developing a future ontological 
	theory, this may be a desirable feature.  
	
	Here, it suffices to point out that a Hamiltonian pertaining to the unitary dynamics generated by permutations will be completely fixed, once an underlying Ising model and its updating rule and time step size are defined. However, with respect to realistic circumstances, trying to pin down a theoretical model based on experimental data, inaccuracies of extracted parameters cannot be avoided. 
	Such unavoidable imprecision will replace the `true' Hamiltonian 
	$\hat H$ -- describing deterministic evolution of ontological states -- by a result based on experiments, 
	$\hat H_{exp}=\hat H+\delta\hat H$. 
	And most likely $\hat H_{exp}$ will generate superposition states, which 
	leads necessarily to a Hilbert space description. This indicates that 
	the classical automaton models considered here are always 
	intrinsically and unavoidably close to becoming quantum 
	mechanical ``by mistake''~\cite{Elze2020}. It suggests that quantum theory 
	more generally might be a {\it {best possible theory}} reflecting our 
	persisting ignorance of the \linebreak ontology beneath. 
	
	
	Ingredients for our present considerations were earlier obtained and 
	reported in references~\cite{Elze2020,Elze2024}. 
	In particular, in~\cite{Elze2024} we presented further 
	motivation and context of attempts to find an ontological 
	basis for quantum theory, including additional references.  
	In order to make this article self-contained, we summarize some of the 
	formal aspects in the Appendix~\ref{appendixa}, while new results are 
	described in the following Section(s). 
	
	It is important to realize that the present study can be considered independently of the background of
	discussions concerning the foundations of quantum theory. In the 
	following derivations, one may consider the Dirac equation as a 
	``classical'' field equation, completely independent of its famous 
	quantum theoretical context. And the question raised and answered 
	affirmatively in this work is can we construct a discrete 
	deterministic cellular automaton that captures the dynamics embodied 
	in this equation?
	
	\section{Results: An Automaton Beneath the Dirac Equation} 
	
	To set the stage we first explain how the {\it {Weyl equation}} 
	for massless spin-1/2 particles is obtained from the 
	cellular automaton described in Appendix~\ref{appendixa.2}, based on the 
	{\it {Cogwheel Model}} reviewed in Appendix~\ref{appendixa.1}. This will be followed by the construction of an automaton representing the Dirac equation for massive spin-1/2 particles in $1+1$ dimensions, which requires additional  structure to incorporate the mass. 
	
	\subsection{The Weyl Equation} 
	\label{subsection2.1}
	In order to arrive at the Weyl equation, a continuum field 
	equation, from the discrete cellular automaton based on two-state Ising spins--equivalently, on block numbers defined in Appendix~\ref{appendixa.2}, Equations~(\ref{A8}) and (\ref{A9}) -- we suitably denote an arbitrary state $|\psi\rangle$ of the spin chain by \clearpage 
	\begin{eqnarray}\label{psinew} 
		|\psi\rangle 
		&:=&|\{ s(2k-1)\},\{ s(2k)\}\rangle ,
		\\[1ex]\label{psinew1} 	
		&=:&|\{ s^L(k)\},\{ s^R(k)\}\rangle 
		, \end{eqnarray} 
	for $s(k):=s_k$, in terms of the spin variable at site $k$, $k=1,\dots,S+1$, and with periodic boundary conditions, $s(2S+1)\equiv s(1)$, $s(2S+2)\equiv s(2)$. Here, 
	we first group the left- and right-moving spin variables 
	together, and then rename them $s^L$ and $s^R$, respectively, and  renumber the even- and odd-numbered sites.  
	
	Considering, for example, the evolution 
	generated by $\hat U$ of Equation~(\ref{A7}) in $l$ steps, the update of a 
	generic state $|\psi_n\rangle$ at time $t=nT$ is exactly described 
	by a pair of equations: 
	\begin{eqnarray}\label{Lmover} 
		s^L_{n+l}(k)-s^L_n(k)&=&
		s^L_n(k+l)-s^L_n(k) , 
		\\[1ex]\label{Rmover} 
		s^R_{n+l}(k)-s^R_n(k)&=&
		s^R_n(k-l)-s^R_n(k) 
		, \end{eqnarray}
	where suitable identical terms were subtracted on both sides 
	of Equations \cite{Elze2020,Elze2024}. 
	
	Then, for $l=1$, Equations (\ref{Lmover}) and (\ref{Rmover}) simply  represent discretized first-order partial differential equations in 1 + 1 dimensions. Remarkably, the left- and right-movers do not interact and up-and-down spins are never flipped here. -- Identical equations hold for the block numbers introduced in 
	Equation~(\ref{A9}) in Appendix~\ref{appendixa.2}. 
	
	Jumping to the continuum limit, we introduce linear combinations $S^\pm :=s^L\pm s^R$ {and} 
	$\bar\sigma^0:=\mathbf{1}_2$, $\bar\sigma^1:=-\sigma_x$ with unit matrix $\mathbf{1}_2$ and Pauli matrix $\sigma_x$. Thus, the \mbox{Equations~(\ref{Lmover}) and (\ref{Rmover})} 
	can be combined and yield the {\it{ left-handed Weyl equation}}: 
	\begin{equation}\label{LWeyl} 
		\bar{\sigma}^\mu\partial_\mu\Psi_L
		=\left (\mathbf{1}_2\partial_t-\sigma_x\partial_x\right )\Psi_L=0 
		, \end{equation}
	for the two-component `spinor' $(\Psi_L)^t:=(S^+,S^-)$---the {\it {right-handed Weyl equation}}, with  
	$\sigma^0:=\mathbf{1}_2$, $\sigma^1:=+\sigma_x$, and $(\Psi_R)^t:=(S^+,S^-)$,  
	\begin{equation}\label{RWeyl} 
		\sigma^\mu\partial_\mu\Psi_R
		=\left (\mathbf{1}_2\partial_t+\sigma_x\partial_x\right )\Psi_R=0 
		, \end{equation}
	is obtained similarly, especially  
	replacing $\partial_x\rightarrow -\partial_x$ in the derivation following \mbox{Equations~(\ref{Lmover}) and (\ref{Rmover}),} which amounts to 
	an additional overall minus sign on the right-hand sides of both equations or exchanging left- and right-movers there. In turn, this is equivalent to numbering the sites of the spin chain from right to left instead of from left to right, as before.  
	
	The continuum limit can be studied in a controlled way, for example, by applying {\it {Sampling Theory}}, 
	similarly as in references~\cite{PRA2014,Shannon,Jerri,Kempf}. 
	This maps the above {\it {finite differences equations}}  one-to-one on continuous space-time partial differential  equations for {\it {bandwidth-limited fields}} $s^L(x,t)$ and $s^R(x,t)$, i.e., continuum equations for functions with an {\it  {ultraviolet cut-off}} corresponding to a nonzero scale  
	$T$. There arise higher-order derivative terms from expanding $\exp (T\partial_t)$ or 
	$\exp (T\partial_x)$ ($c=1)$, which disappear for $T\rightarrow 0$. -- Note that on discrete 
	lattice points, the functions $s^{L,R}(x,t)$ can only assume the Ising spin values $s_k=\pm 1$, while they can be real-valued in between. This limitation is overcome by assuming block numbers as underlying discrete variables which, by construction, range over a larger set of non-negative integers; see Appendix~\ref{appendixa.2}. 
	
	Obviously, there are no mass terms in the Weyl equation. However, 
	in reference~\cite{Elze2024}, we showed that the implied signal velocity can be changed to differ from $c=1$ by 
	modifying the update rule in simple ways. Nevertheless, this  
	does not correspond to introducing a mass, as will be needed for 
	the Dirac equation, cf. the following {Section\,2.2. } 

	\subsection{Reverse Engineering an Automaton Based on Permutations from the Dirac Equation} 
	
	We have studied the deterministic cellular automaton based on an Ising 
	spin chain with exchange interactions, i.e., transpositions of two-state variables. This has led to the Weyl equation for {\it {massless}} 
	particles in the continuum limit, but it did not provide 
	insight into how to generate the mass terms for the Dirac equation. In order to 
	proceed, we elaborate in the following the {\it {``reverse engineering''}} approach suggested previously~\cite{Elze2024}.    
	
	Consider the real representation of the Dirac equation in 1 + 1 dimensions 
	\cite{KauffmanNoyes}: 
	\begin{equation}\label{KauffmanNoyesI}
		\partial_t\Psi =
		\left (\sigma_z\partial_x-i\mu\sigma_y\right )\Psi 
		, \end{equation}
	where 
	$\Psi^t:=(\Psi_1,\Psi_2)$ is the two-component real spinor field, $\sigma_{y,z}$ are the usual Pauli matrices, and $\mu$ denotes a dimensionless mass parameter; here, we choose units such that $c=\hbar/m=1$ for the dimensional mass scale $m$. -- The Equation (\ref{KauffmanNoyesI}) is equivalent to 
	two coupled partial differential equations:  
	\begin{eqnarray}\label{Psi1} 
		\partial_t\Psi_1&=&\partial_x\Psi_1-\mu\Psi_2
		, \\ [1ex] \label{Psi2} 
		\partial_t\Psi_2&=&-\partial_x\Psi_2+\mu\Psi_1 
		. \end{eqnarray} 
	{Our} 
	strategy now is to discretize them, going backward in the direction of Equations~(\ref{psinew})--(\ref{Rmover}), and see the changes brought about by the presence of mass terms. 
	
	We replace $\Psi_1(x,t)\rightarrow s_n^L(k)$ and 
	$\Psi_2(x,t)\rightarrow s_n^R(k)$, i.e., directly in terms of what have been left- and right-movers, respectively. 
	Thus, 
	instead of Equations~(\ref{Lmover}) and (\ref{Rmover}), we \linebreak obtain 
	here 
	\begin{eqnarray}\label{Lmoverm} 
		s^L_{n+1}(2k-1)&=&
		s^L_n(2k+1)-\mu s_n^R(2k) , 
		\\[1ex]\label{Rmoverm} 
		s^R_{n+1}(2k)&=&
		s^R_n(2k-2)+\mu s_n^L(2k-1)  
		, \end{eqnarray}
	for $k=1,\dots,S+1$, with periodic boundary conditions, and canceling identical terms on both sides of the equations that arise from the discretization of the derivatives. 
	Herein, the notation is as earlier in Equation~(\ref{psinew}), 
	yet we left the superscripts $\phantom .^{L,R}$ for clarity: the left-movers (right-movers) occupy the odd (even)-numbered sites of a chain, cf. Figure\,\ref{fig2} (top). 
	
	{From the} {Equations~(\ref{Lmoverm}) and (\ref{Rmoverm})}, {one can derive second-order finite difference equations. They agree with those obtained directly by discretizing the second-order wave equations for the spinor components in the continuum.} {Equations (\ref{Lmoverm}) and (\ref{Rmoverm})} {can also be inverted in order to obtain the automaton equations for the deterministic evolution backwards, determining the earlier $s^{L,R}_n$ in terms of the later $s^{L,R}_{n+1}$.} 
		
	Note that the mass terms add simple inhomogeneities to the previous equations for left- and right-movers. Therefore, the variables can no longer be related to {\it {two-state}} Ising spins as before, cf. {Section \ref{subsection2.1}.} 
	However, we set 
	$\mu =1$ henceforth, such that they may still be considered integer valued. -- This could be consistent with an underlying 
	Ising spin chain if we repeatedly perform the transformation 
	on Equations (\ref{Lmoverm}) and (\ref{Rmoverm}) that we discuss in Appendix~\ref{appendixa.2}, following Equations~(\ref{A8}) and (\ref{A9}). It would lead to 
	possibly large {\it {block variables}} instead of the 
	{\it {two-state}} Ising spins. However, it turns out that the attempt to construct an automaton in the way sketched in~\cite{Elze2024} 
	does not succeed and the set-up has to be changed more profoundly. This 
	will be shown in the following {Section \ref{subsection2.2.1}}. 

	\subsubsection{Doing Arithmetic by Permutations} 
	\label{subsection2.2.1}
	Our aim is to handle also the additive contributions produced by the mass terms by {\it {permutations}}, 
	besides those that produce the left- and right-moving kinematics in the above equations, as we have seen earlier. In this way, the update rule of the new cellular automaton will altogether be described by permutations, extending the {\it {Cellular Automaton Interpretation}} from the case of the Weyl equation {(Section\,2.1)} 
	to the Dirac equation. 
	
	However, as indicated above, we need to generalize the 
	model of the Ising spin chain in order to accommodate 
	the arithmetic implied by the mass terms of the Dirac 
	equation.  
	Instead of an Ising spin chain with periodic boundary conditions, we will implement a {\it {``Necklace of Necklaces''}}. This we already alluded to in 
	reference~\cite{Elze2024}. However, its torus-like 
	topology will now be made use of differently, once again based on {\it {two-state Ising spins}}.  
	
	Borrowing notation from quantum theory, the state of the automaton at time t = nT is represented by
	\begin{equation}\label{PsiDirac} 
		|\Psi\rangle_n:=|s_{1,\; l_{1}},\;\dots,\; 
		s_{2k,\; l_{2k}},\;\dots,\; 
		s_{2k+1,\; l_{2k+1}},\;\dots,\;  
		s_{2S,\; l_{2S}}\rangle 
		,\end{equation} 
	where the spin variables can assume values $s=\pm 1$. 
	We note that the variables here are labeled by {\it {two indices}}, 
	an either even or odd index for the ``longitudinal'' position, 
	with $k=1,\dots ,S\;$, and an index for the ``transverse'' position, 
	$-S,-S+1,\dots ,l_k,\dots,S-1,S$; the index $k$ on $l_k$ will mostly be suppressed when no confusion arises. 
	
	Furthermore, the following torus-like {\it {periodic boundary conditions}} are assumed: 
	\begin{eqnarray}\label{kperiod}
		&\;&s_{2S+1,\; l}=s_{1,\; l},\;\; 
		s_{2S+2,\; l}=s_{2,\;  l},\;\;\mbox{for all}\; l 
		, \\[1ex] \label{lperiod}
		&\;&\;s_{k,\; S+1}=s_{k,\; -S},\;\; 
		s_{k,\; -S-1}=s_{k,\;  S},\;\;\mbox{for all}\; k 
		. \end{eqnarray} 
	Thus, the index $k$ labels even- and odd-numbered sites along a 
	{\it {``necklace''}} in the longitudinal direction and  
	each such site is part of a transverse {\it {``necklace''}}; altogether sites are labelled by $k,l_k$, often simplified to $k,l$, hence the name {\it {``Necklace of Necklaces''}}. 
	
	The states of the automaton, cf. Equation~(\ref{PsiDirac}), are further restricted to have, for each longitudinal position, say $k'$, {\it {exactly one spin up}} with 
	$s_{k',\; l'}=+1$ and {\it {all other spins down}} with value $s_{k',\; l\neq l'}=-1$. 
	
	The operator $\hat L(k')$ which measures the position of the {\it {one up spin}} at a  given $k'$ is then defined by
	\begin{equation}\label{Lk}
		\hat L(k'):=\sum_{l=-S}^S 
		\frac{l}{2}(\hat\sigma^z_{k',\; l}+\mathbf{1}_2) 
		, \end{equation}
	where $\hat\sigma^z_{k',\; l}$ denotes the usual diagonal 
	Pauli matrix acting on the spin 
	$s_{k',\; l}$. 
	The restricted automaton states are eigenstates 
	of this operator, such that 
	for $s_{k',\; l'}=+1,\; s_{k',\; l\neq l'}=-1$ we obtain  
	$\hat L(k')|\Psi\rangle_n=l'|\Psi\rangle_n$\,. 
	
	We shall now employ the transverse position $l'$ of a single up spin with given longitudinal position $k'$ to encode the integer value of the variables  $s_n^L(k')$ ($k'$ odd) or $s_n^R(k')$ ($k'$ even) that are described by the  Equations~(\ref{Lmoverm}) and (\ref{Rmoverm}). Additions and subtractions due to the mass terms then correspond to 
	{\it {shifts of up spin positions}}. 
	
	A {\it {positive unit shift}} for a state with a single up spin at $k',l'$ is effected by moving 
	{\it {all}} spins with this $k'$ one step towards increasing transverse position $l_{k'}$. This is achieved by a complete permutation -- see Appendix~\ref{appendixa.1} -- with 
	a {\it {unitary operator}} $\hat U_{\perp}$:  
	\begin{equation}\label{Utransverse}
		\hat U_{\perp}(k')|\dots,\; 
		s_{k',\; l_{k'}},\;\dots\rangle 
		:=|\;\dots,\; 
		s_{k',\; l_{k'}+1},\;\dots\rangle 
		,\end{equation}
	for all spins at fixed $k'$ with $l\in [-S,S]$, respecting the periodic boundary condition.  
	This operator is explicitly given in terms of {\it {pairwise transpositions}} of nearest neighbor spins: 
	\begin{equation}\label{Utransverse1} 
		\hat U_{\perp}(k'):=\prod_{l=-S}^S\hat P_{l\;l+1}(k')  
		,\end{equation} 
	which are discussed in Appendix~\ref{appendixa.2}. Using unitarity, 
	with $\hat U^{-1}_\perp =\hat U^\dagger_\perp$, a {\it {negative unit shift}} can likewise be expressed in terms of elementary transpositions.  
	
	It is now clear that additions and subtractions required by the finite difference equations of motion 
	can indeed be performed by suitable (multiple) permutations or transpositions of the encoding Ising spins.   
	
	We proceed in several steps. -- First, we re-express the right-hand sides 
	of \linebreak Equations~(\ref{Lmoverm}) and (\ref{Rmoverm}) by corresponding terms using the above restricted Ising spin states and permutations: 
	\begin{eqnarray}\label{rhsLmoverm}
		s^L_n(2k+1)-s_n^R(2k)
		&\rightarrow&
		\left (\hat U_{\perp}(2k+1)\right )^{-s_n^R(2k)}|\Psi\rangle_n 
		, \\[1ex]\label{rhsRmoverm} 
		s^R_n(2k)+s_n^L(2k+1) 
		&\rightarrow&
		\left (\hat U_{\perp}(2k)\right )^{+s_n^L(2k+1)}|\Psi\rangle_n   
		, \end{eqnarray} 
	where we shifted $k\rightarrow k+1$ in (\ref{rhsRmoverm}), as compared to Equation~(\ref{Rmoverm}) (with $\mu =1$). 
	The exponents on the right-hand sides of (\ref{rhsLmoverm}) and 
	(\ref{rhsRmoverm}) 
	$\hat L(k')$ defined in Equation~(\ref{Lk}): 
	\begin{eqnarray}\label{L2k}
		s_n^R(2k)|\Psi\rangle_n&=&\hat L(2k)|\Psi\rangle_n
		, \\[1ex]\label{L2k+1}
		s_n^L(2k+1)|\Psi\rangle_n&=&\hat L(2k+1)|\Psi\rangle_n
		.  
	\end{eqnarray} 
	The updates described in (\ref{rhsLmoverm})--(\ref{L2k+1}), 
	so far, {\it {cannot}} be performed {\it {consecutively}}. This 
	would obviously violate the evolution generated by the 
	right-hand sides of \mbox{Equations~(\ref{Lmoverm}) and (\ref{Rmoverm}). }
	
	Instead, they have to take place simultaneously  
	for all positions $k$, mediated locally by a 
	{\it {scattering operator}} $\hat {\cal M}(k)$, which is introduced in
	\begin{equation}\label{scatteringopk}
		|\Psi\rangle_n'=\prod_{k=1}^{S}\hat {\cal M}(k)|\Psi\rangle_n\; :=\;\prod_{k=1}^{S}\;\;\left ] 
		\begin{array}{c} 
			\left(\hat U_{\perp}(2k)\right )^{+\hat L(2k+1)} \\ 
			\left(\hat U_{\perp}(2k+1)\right )^{-\hat L(2k)} 
		\end{array} 
		\right [\; |\Psi\rangle_n 
		. 
	\end{equation}
	The inverted brackets {$]\dots [$} 
	are meant to indicate that the 
	operations listed in between act on {\it {two neighboring transverse 
			necklaces}} situated at longitudinal position $2k$ and $2k+1$, respectively, 
	and yield these two necklaces updated according to the right-hand sides of 
	(\ref{rhsLmoverm})  and (\ref{rhsRmoverm}) -- Examples illustrating this process are provided 
	in Appendix~\ref{appendixa.3}. -- Since operations for one 
	such even--odd position nearest neighbor pair do 
	not interfere with those for another one, the product for all pairs is 
	taken in the end. 
	
	We may rewrite the complete permutations implemented by unitaries in Equation~(\ref{scatteringopk}) as exponentials of the corresponding Hamiltonian, 
	$\hat U_\perp (k') =\exp (-iT\hat H_\perp (k'))$, which has been explicitly given in Appendix~\ref{appendixa.1}; here, it acts on all spins with fixed longitudinal position $k'$ and with $2S+1$ varying indices for the related transverse necklace. Since, e.g., 
	$\hat H^\perp (2k)$ and $\hat L(2k+1)$ act on spins with different longitudinal index, they commute and can be exponentiated side by side; also, $[\hat H^\perp (2k+1),\hat L(2k)]=0$. Thus, we obtain 
	\begin{equation}\label{scatteringopk1}
		\prod_{k=1}^{S}\hat {\cal M}(k)\; =\;
		\left ] 
		\begin{array}{c} 
			\exp\left (-iT\sum_{k=1}^S\hat H_{\perp}(2k)\hat L(2k+1)\right ) \\ 
			\exp\left (+iT\sum_{k=1}^S\hat H_{\perp}(2k+1)\hat L(2k)\right ) 
		\end{array} 
		\right [
		,  
	\end{equation}
	which represents the update operations to be applied on 
	the automaton state $|\Psi\rangle_n$ to yield the intermediate results corresponding to the right-hand sides of  Equations~(\ref{Lmoverm}) and (\ref{Rmoverm}). 
	
	Finally, in order to complete the evolution equation for the automaton, we have to take into account that the intermediate 
	state $|\Psi\rangle_n'$ of Equation~(\ref{scatteringopk}) still has to be modified such that the {\it {left- and right moving  kinematics}} is incorporated.  
	
	According to Equation~(\ref{Lmoverm}), the 
	result produced by the right-hand side of 
	this equation at longitudinal position 
	$2k+1$, especially in absence of a mass term (cf. ``Weyl automaton''), 
	has to move {\it {left}} by two positions to $2k-1$, for all $k$. Similarly, according to Equation~(\ref{Rmoverm}), the intermediate result at even-numbered longitudinal position $2k$ has to move right by two positions to $2k+2$, for all $k$, cf. Figure\,\ref{fig2} (top). These moves complete the 
	{\it {one-step~update}}: 
	\begin{equation}\label{onestepup}
		|\Psi\rangle_n\; \stackrel{|\Psi\rangle_n')}{\longrightarrow}\; |\Psi\rangle_{n+1}
		,\end{equation} 
	related to the time interval $T$, as before. See also simplest examples in Appendix~\ref{appendixa.3}.
	
	From the discussion in Appendix~\ref{appendixa.2}, we know how left- and right-movers 
	are generated by transpositions acting on a spin chain. Presently, however, we have to move the complete transverse necklaces of spins. This 
	is effected by an obvious generalization of the previous unitary of  Equation~(\ref{A7}), i.e., by a unitary operator $\hat U_{kin}$ defined as: 
	\begin{equation}\label{Ukin}
		\hat U_{kin}:=\prod_{l=-S}^S
		\prod_{k=1}^S\hat P_{2k-1\;2k;\;l}\prod_{k'=1}^{S}\hat P_{2k'\;2k'+1;\;l}
		=:\exp (-iT\hat H_{kin}) 
		,\end{equation} 
	where the transpositions $\hat P_{ij;\;l}$ exchange the nearest neighbor spins 
	at longitudinal positions $i$ and $j$, 
	respectively, with identical transverse 
	index $l$. Similarly as before, the corresponding Hamiltonian 
	$\hat H_{kin}$ can be given analytically as a polynomial in  $\hat U_{kin}$ and $\hat U_{kin}^\dagger$~\cite{Elze2020}. 
	
	Thus, we finally arrive at the {\it {one-step update rule}} for the cellular automaton reverse-engineered from the Dirac equation in 1 + 1 dimensions: 
	\begin{equation}\label{DiracCA}  
		|\Psi\rangle_{n+1}=\hat U_{kin}\prod_{k=1}^{S}\hat {\cal M}(k) 
		|\Psi\rangle_n
		. 
	\end{equation} 
	This presents our main result. -- Following the arguments presented 
	in this work, the \linebreak Equation~(\ref{DiracCA}) summarizes the 
	behavior of a new {\it {deterministic cellular automaton}} for the 
	Dirac equation in $1+1$ dimensions. Emphasis lies on the deterministic evolution, while earlier attempts have been probabilistic or fully quantum mechanical instead. 
	
	\section{Discussion} 
	
	We emphasize that the dynamics of the {\it {``Dirac automaton''}} described 
	above is completely based on permutations of underlying Ising spins, 
	the two-state variables that define its ontological states, in 
	accordance with the {\it {Cellular Automaton Interpretation of Quantum Mechanics}}~\cite{tHooftBook}. 
	
	We had to introduce the {\it {scattering operator}} 
	$\hat {\cal M}(k)$, Equation~(\ref{scatteringopk}), in order to encode the 
	addition/subtraction arising on the right-hand side of the 
	Dirac equation through the mass term, cf.  Equations~(\ref{KauffmanNoyesI})--(\ref{Psi2}), in a deterministic way through 
	suitable permutations. Our finding in this  
	{\it {``Necklace of Necklaces''}} model has been that permutations of spins 
	on one necklace are conditioned by a measurement result on the neighboring 
	one. -- It appears that all earlier attempts to formulate 
	an automaton for the Dirac equation, to the best of our knowledge, instead 
	had to make use of interfering amplitudes borrowed from quantum mechanics or of probabilistic evolution~\cite{Feynman,Iwo,DArianoCA,Margolus,Wetterich22,Ord,QCA24}. In distinction  from our classical automaton, these models are cellular automata of a different (nondeterministic) kind. They are known as {\it {quantum cellular automata}} and {\it {probabilistic cellular automata}}, respectively. See also  references 
	\cite{Khrennikov,AtmanspacherPrimas,Khrennikov1,Khrennikov2,Scholes24,Wetterich24} for related more general discussions of probability and quantum~theory. 
	
	This may all seem rather tedious. However, it illustrates that addition 
	and subtraction on a finite set of adjacent 
	integers with periodic boundary conditions can be obtained simply from permutations of these numbers, which represent discrete 
	automaton states. Such a description is obviously not limited to small sets of such  states. 
	
	Clearly, since there are non-commuting transpositions (exchange interactions) $\hat P$ and operators $\hat L$ involved which measure the transverse 
	position of an up spin, it will not be straightforward to 
	extract the overall Hamiltonian governing this automaton. An 
	approximate result can always be obtained with the help of a  
	Baker--Campbell--Hausdorff formula; yet, an exact result 
	would be of interest and we intend to report this elsewhere.  
	
	In the end, the discrete integer variables that we have dealt with can be mapped to continuous real ones by invoking {\it {Sampling Theory}}, as we recalled in {Section \ref{subsection2.1};} 
	\linebreak see~\cite{PRA2014,Shannon,Jerri,Kempf}. We remind  the reader that the value of the wave function in the discrete model is  determined at each longitudinal spatial position $2k$ or $2k+1$ 
	by \mbox{Equations~(\ref{L2k}) and (\ref{L2k+1}).} In this way, letting the necklaces of the model become sufficiently large -- i.e., the {\it {separately conserved numbers of up and down Ising spins}} -- the spinor field is free to move so 
	that its absolute value can possibly become large in some places or small in others in the \linebreak continuum limit.  
	
	\section{Conclusions}
	
	To summarize, a discrete deterministic automaton representation for the 
	Dirac equation in 1 + 1 dimensions is presented, which is based on  Equations~(\ref{Lmoverm}) and (\ref{Rmoverm}). In order to incorporate the 
	structure of these finite differences equations, especially the 
	arithmetic required by their right-hand sides, by making use of permutations 
	of two-state spins in an underlying classical Ising model, 
	a torus-like compactification of the state space has been introduced. Here,  
	the real discrete value of the Dirac field is encoded in the position of an  
	up spin on a transverse ``necklace'' together with its longitudinal 
	spatial position on a chain with periodic boundary conditions---altogether forming a {\it {``Necklace of Necklaces''}}. 
	
	Generalizing an earlier study of the Weyl equation \cite{Elze2024}, which  features no mass term, the evolution of this automaton is determined by  elementary spin exchange operations, building up permutations from 
	transpositions. This is in accordance with the {\it {Cellular Automaton Interpretation of QM}} which requires to build discrete deterministic models  with dynamics generated by {\it {permutations of ontological states}} 
	\cite{tHooftBook}. Thus, the {\it {``Necklace of Necklaces''}} model 
	described in this work {\it {never}} leads to quantum mechanical superposition 
	states, it is a classical automaton. 
	
	However, we recall that quantum 
	mechanical features enter such models ``by mistake'' or, rather,  
	mathematical convenience, as discussed extensively earlier 
	\cite{tHooftBook,Elze2020,Elze2024,Wetterich22,Wetterich24}. 
	Instead of resuming this here, we conclude by listing 
	several conceivable extensions of \linebreak the model:    
	\begin{itemize}
		\item	The generalization to more than $1+1$ dimensions should be feasible, 
		notwithstanding some necessary bookkeeping efforts.   
		\item	The effective Hamiltonian governing the discrete model should be derived and the continuum limit be understood. 
		\item Symmetries and conservation laws of cellular automata are of 
		general interest beyond the present context.
		\item   A field theory of fermions obeying the Pauli principle should be formulated (relating two-state Ising spins to occupation numbers?).  
		\item  An external scalar potential can be considered as a spatially 
		varying mass, yet interactions with dynamical gauge fields will require 
		additional considerations. 
	\end{itemize} 
	
	This may show that there is ample room for future work. 
	
	\vspace{6pt} 
	
	\funding{This research received no external funding.}
	
	
	\dataavailability{{The original contributions presented in this study are included in the article. Further inquiries can be directed to the corresponding author.}} 
	
	\acknowledgments{It is a pleasure to thank Ken Konishi, Gregory 
		Scholes, and Andrei Khrennikov for discussions.}
	
	\conflictsofinterest{The author declares no conflicts of interest.}

		
	
	\appendixtitles{no} 
	\appendixstart

	\appendix 
	
	\section[\appendixname~\thesection]{}\label{appendixa}
	\appendixtitles{yes} 
	\subsection[\appendixname~\thesection]{Permutations and the Cogwheel Model} \label{appendixa.1}
	
	The {\it {Cellular Automaton Interpretation of QM}} suggests to build 
	discrete deterministic models with dynamics generated by 
	{\it {permutations of ontological states}}~\cite{tHooftBook}. 
	
	The periodicity of such models, without fusion or fission of trajectories in state space, assimilates them to QM models. 
	
	Consider $N$ objects, $A_1,A_2,\dots ,A_N$ (``states''), which are mapped in 
	$N$ steps onto one another, involving {\it {all}} states. This can be represented 
	by a {\it {unitary}} {$N \times N$} 
	matrix with {\it {one}} off-diagonal phase per column and row and with zeroes elsewhere: 
	\begin{equation}\label{A1} 			
		\hat U_N:= 
		\left (
		\begin{array}{c c c c c} 
			0           & .           & .     & 0       & e^{i\phi_N} \\ 
			e^{i\phi_1} &0            &       & .       & 0  \\ 
			0           & e^{i\phi_2} & .     &         & .  \\
			.           &             & .     & .       & .  \\
			0           & .           & 0     & \;e^{i\phi_{N-1}} & 0  \\
		\end{array}
		\right )
		.\end{equation} 
	The basis of vectors, on which such a {\it complete permutation} $\hat U_N$  acts, we refer to as the {\it {standard basis}}. 
	
	Since $(\hat U_N)^N=e^{i\sum_{k=1}^N \phi_k}\;\mathbf{1}$, the eigenvalues of 
	the corresponding {\it {Hamiltonian}} $\hat H_N$, which is defined {\it {via}} 
	$\hat U_N=:e^{-i\hat H_NT}$, are obvious. Here, $T$ serves to introduce a 
	time scale; its inverse provides an energy scale for the Hamiltonian. (We always use units such that $\hbar =c=1$.) 
	Thus, we may write down the matrix elements of the Hamiltonian ($n=1, \ldots, N$): 
	\begin{equation}\label{A2}
		(\hat H_N)_{nn}=\mbox{diag}\Big (\frac{1}{NT}\big (2\pi (n-1)
		-\sum_{k=1}^N\phi_k\big )\Big ) 
		,\end{equation} 
	which refer to {\it {diagonal(izing) basis}}. Since the arbitrary phases $\phi_k$ 
	play no role in what follows, they will be omitted. 
	
	The diagonal and standard bases are unitarily related by discrete  Fourier transformations. This allows us to calculate the matrix elements  of the Hamiltonian with respect to the standard basis: 
	\begin{equation}\label{A3}  
		(\hat H_N)_{nn}=\frac{\pi}{NT}(N-1)
		,\;\; n=1,\dots ,N 
		,\end{equation} 
	and  
	\begin{equation}\label{A4} 
		(\hat H_N)_{n\neq m}=\frac{\pi}{NT}
		\Big (-1+i\cot\big (\frac{\pi}{N}(n-m)\big ) \Big ) 
		,\;\; n,m=1,\dots ,N  
		,\end{equation} 
	and $(\hat H_N)_{nm}=(\hat H_N)^*_{mn}$, as it should be. 
	
\begin{figure} [H]
		
		\includegraphics[width=0.7\columnwidth 
		]{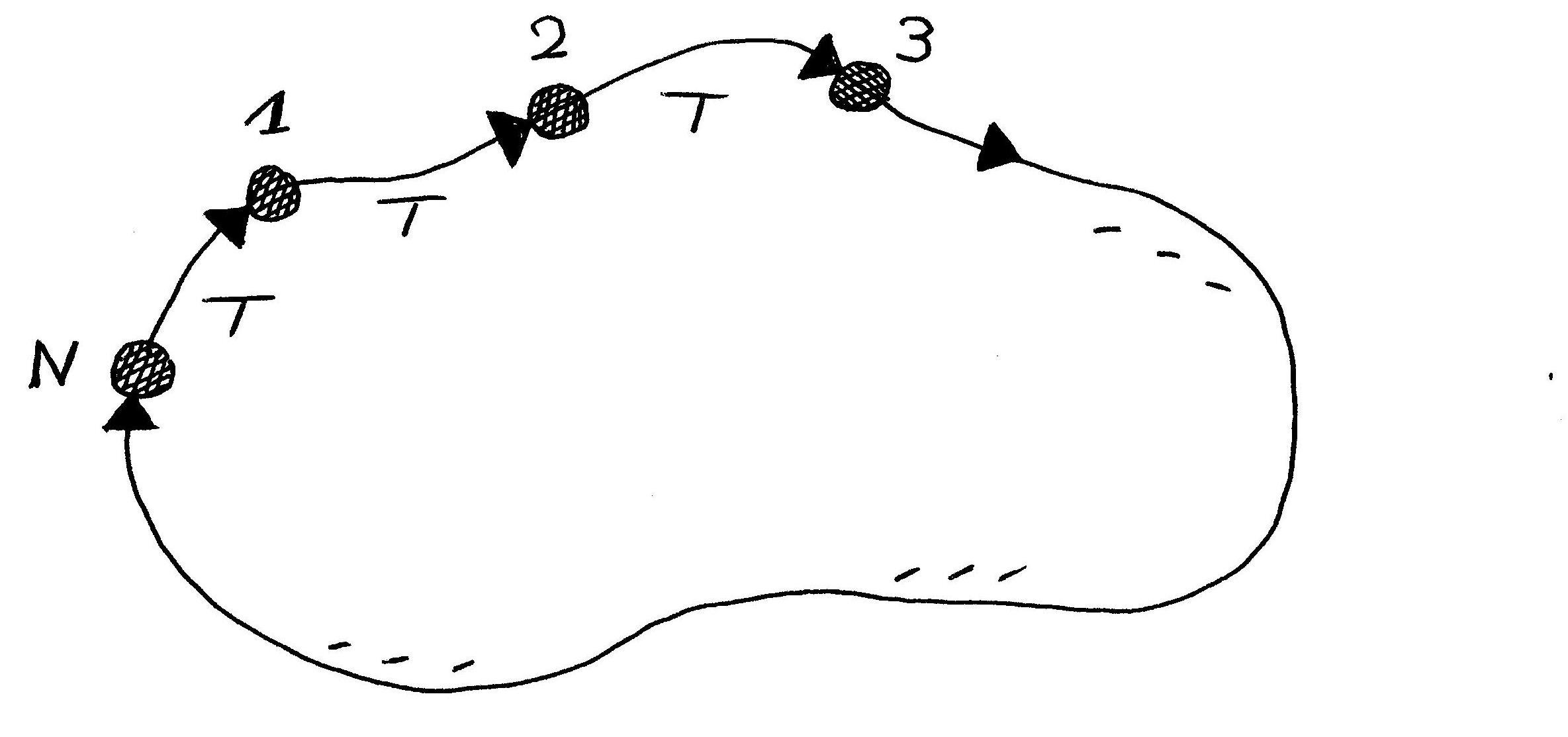} 
		
		\caption{{A } 
			{\it {Cogwheel}} jumps periodically through $N$ states, taking time $T$ for each jump. {\it {From}}~\cite{ElzehybridCA}. 
			\label{fig1}} 
	\end{figure}

	What we have represented here is known as the {\it {Cogwheel Model}}; 
	see~\cite{tHooftBook,Elze2020} for further details and additional references. 
	With $N\rightarrow\infty $, $T\rightarrow 0$, $\omega :=1/NT$ fixed, the {\it {Cogwheel Model}} can be mapped onto a description of the quantum harmonic oscillator. Such a simple automaton is illustrated in Figure~\ref{fig1}. 
	
	\subsection[\appendixname~\thesection]{A Chain of Ising Spins with Exchange Interactions} \label{appendixa.2}
	
	We consider a chain of $2S$ classical two-state spins with  periodic boundary conditions, which will be always in one of $2^{2S}$ possible {\it {ontological states}}. The latter are specified by the two-state variables $s_k=\pm 1,\;\;k=1,\dots ,2S,2S+1$, with $s_{2S+1}\equiv s_1$. 
	
	Interactions are defined as transpositions of 
	the two-state variables ({\it {spin exchange}}) of nearest neighbors: 
	\begin{equation}\label{A5}
		\hat P_{ij}|s_i,s_j\rangle :=|s_j,s_i\rangle ,\;\;  
		\hat P_{2S\;2S+1}\equiv\hat P_{2S\;1} 
		, \end{equation}
	with the following properties of the exchange operators: 
	\begin{equation}\label{A6}
		\hat P^2=\mathbf{1},\;   
		[\hat P_{ij},\hat P_{jk}]\neq 0 ,\;\;  
		\hat P_{ij}= (\underline{\hat\sigma}_i\cdot\underline{\hat\sigma}_j+\mathbf{1})/2  
		, \end{equation}
	where the last equation recalls the relation between   
	exchange operators and vectors formed by the three Pauli matrices, 
	e.g., $\underline{\hat\sigma}_k$ acting on the spin at site  $k$; a generalization for larger than two-state spins exists as well. 
	
	The dynamics of the chain are then defined by permutations of 
	the chain states which are generated by multiple applications of 
	the exchange operators:  
	\begin{equation}\label{A7}
		\hat U:=\prod_{k=1}^S\hat P_{2k-1\;2k}\prod_{k'=1}^{S}\hat P_{2k'\;2k'+1}
		=:\exp (-i\hat HT) 
		,\end{equation} 
	indicating once more the definition of the corresponding Hamiltonian. The latter can be obtained analytically as a polynomial in $\hat U$ and $\hat U^\dagger$~\cite{Elze2020}. 
	
	We note that, reading the product of transpositions 
	in Equation~(\ref{A7}) from the right to the left, first, all even transpositions are performed and then all odd ones, which commute 
	among each other, respectively---here, even/odd refers to whether 
	the first index $k$ in $\hat P_{k<l}$ is even/odd. This separation of the even/odd transpositions implements a finite 
	signal velocity in the model, such that a perturbation of a particular spin reaches only its nearest neighbors in one update step, which is effected by applying the unitary $\hat U$ once on the state of the chain. 
	
\begin{figure} [H]
		
		\includegraphics[width=0.99\columnwidth 
		]{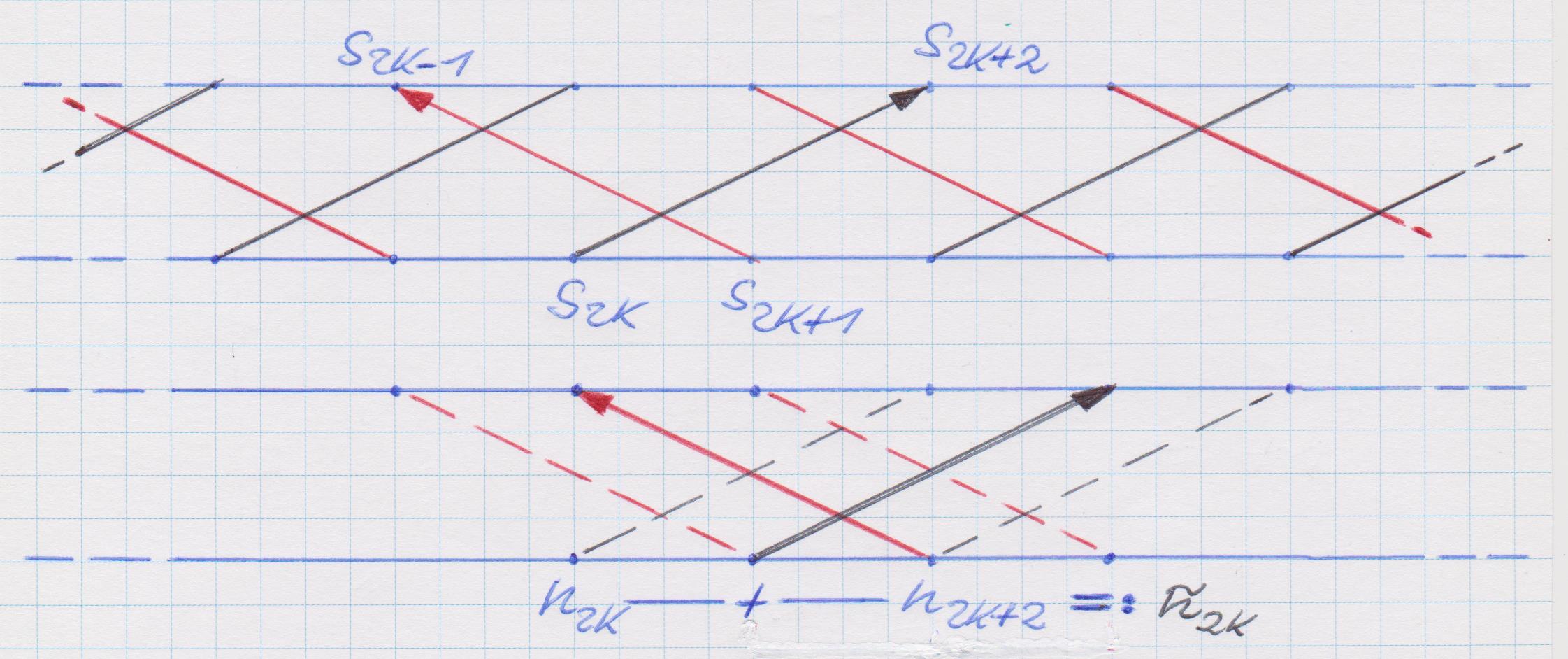}   
		
		\caption{({\bf Top}): {Spin} 
			variables on {\it {odd/even}} sites are transported to the next {\it {odd/even}} sites to the {\it {left/right}}, respectively, in one 
			update by $\hat U$, Equation~(\ref{A7}), of the spin chain automaton. ({\bf Bottom}): Occupation numbers on neighboring even sites are added to form right-moving block variables, $n_{2k}+n_{2k+2}=:\tilde n_{2k}$; likewise, left-moving ones, $n_{2k+1}+n_{2k+3}=:\tilde n_{2k+1}$. They are illustrated by arrows displaced by one unit to the right for better visibility.
			\label{fig2}} 
	\end{figure}

	The evolution of an arbitrary initial state of the spin chain is very simple and obeys a number of conservation laws~\cite{Elze2020}. Writing down explicitly a few update steps for 
	a short chain, it can be seen as a composition of {\it {Cogwheel}} like motions of {\it {left-movers}} and {\it {right-movers}}, 
	concerning the spin variables on the {\it {odd}} and {\it {even}} chain sites, respectively. Recalling the periodic boundary condition on the chain, we obtain $(\hat U)^S=\mathbf{1}$ for a chain of $2S$ spins. The  
	behavior of this cellular automaton is illustrated in Figure\,\ref{fig2} (top).  
	
	
	In order to arrive at a continuum theory of the automaton which leads to the Weyl equation, we need to have access to the arithmetic of variables that are allowed to assume sufficiently large or small integer values, rather than the elementary two-state spins with $s_k=\pm 1$ 
	employed so far. It is useful to introduce site {\it {occupation numbers}}: 
	\begin{equation}\label{A8}
		n_{k}:=(s_k+1)/2\; (=0,1)
		.\end{equation} 
	These can be thought of as eigenvalues of a 
	fermionic number operator $\hat n(k)$, which might be of interest later. They preserve the feature of variables associated with even/odd sites to be right-moving/left-moving, respectively, under the updating affected by $\hat U$. Furthermore, they inherit the periodic boundary condition. 
	
	Next, we perform an {\it {invertible linear transformation}} on the set of occupation numbers defined by: 
	\begin{equation}\label{A9}
		\tilde n_{k}:=n_{k}+n_{k+2},\;k=1,\dots ,2s  
		,\end{equation} 
	cf. Figure\,\ref{fig2} (bottom).  
	Iterating this transformation a sufficiently large number of 
	$r$ times we arrive at {\it {block variables}} $\tilde n^{(r)}$  that range between zero and $2^r$. Again, they move left/right under $\hat U$ from an odd/even site, with periodic boundary 
	conditions. 
	
	We remark that the block variables must be distinguished from 
	coarse-grained variables that would be obtained by 
	some decimation of the number of degrees of freedom.  
	
	\subsection[\appendixname~\thesection]{Illustration of Scattering Operator $\hat {\cal M}$ and Automaton Update} \label{appendixa.3}
	
\begin{figure} [H]
		
		\includegraphics[width=0.99\columnwidth 
		]{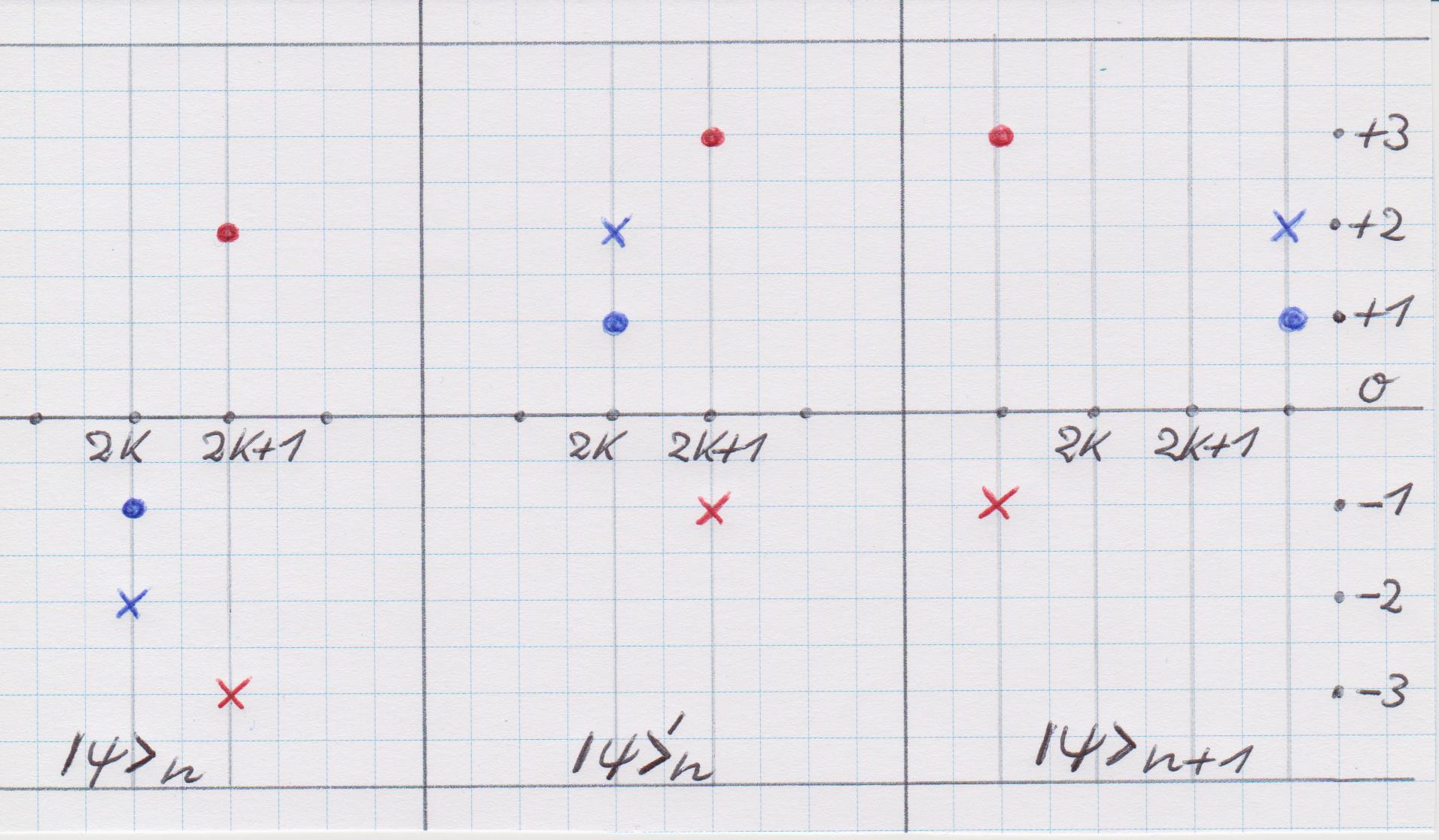}   
		
		\caption{({\bf Left}): {Two} 
			initial conditions (heavy dots, crosses) for $|\Psi\rangle_n$ on two neighboring 
			transverse necklaces attached to the even and odd longitudinal sites $2k$ (blue) and $2k+1$ (red), respectively. 
			Only the transverse positions (drawn vertically, ranging in $[-3,+3]$) of 
			the single up spins are marked; all others are down. ({\bf Middle}): Update of the initial state by the scattering operator 
			$\hat {\cal M}(k)$, Equation~(\ref{scatteringopk}),  
			gives the intermediate result $|\Psi\rangle_n'$. Note that the blue cross 
			has felt the periodic boundary condition 
			in the vertical direction, according to $-2+(-3)\equiv +2$. ({\bf Right}): To obtain the final result of the one-step update, $|\Psi\rangle_{n+1}$, the up spins at even and odd 
			longitudinal positions have to move two positions to the right and left, respectively. 
			\label{fig3}} 
	\end{figure}

	In Figure\,\ref{fig3} we illustrate by two examples how the {\it scattering operator} 
	$\hat {\cal M}(k)$, defined in  Equation~(\ref{scatteringopk}), acts in the update process of the automaton. 
	
	We have shown here an automaton that is very small (with periodic boundary conditions) in the transverse direction for simplicity. 
	Eventually, a numerical simulation of consecutive updates would be amusing 
	to see for a large automaton---one that follows the deterministic rules constructed from the Dirac equation in this work. 
	
	\begin{adjustwidth}{-\extralength}{0cm}
		
		\printendnotes[custom] 
		
		\reftitle{References}
		
		

		\PublishersNote{}
	\end{adjustwidth}
\end{document}